\documentclass{ws-procs9x6}

\begin{document}
\title{Probing the Short-Range Dynamics in Exclusive Scattering 
off Polarized Deuteron}
\author{Misak~M.~Sargsian}

\address{Department of Physics,  Florida International University, 
Miami, FL 33199 }  


\maketitle

\abstracts{Employing the polarization degrees of freedom in the deuteron 
allows to isolate smaller than average inter-nucleon distances in the 
deuteron. As a result one can identify set of high $Q^2$ reactions 
off polarized deuteron which are particularly sensitive to 
the short range dynamics of strong interaction. We concentrate on 
the studies of several aspects of the short range phenomena. These are 
the relativistic dynamics of electron--bound-nucleon scattering, color 
coherence in high $Q^2$ electro-production as well as the formation of the 
vector mesons in coherent electro-production from the deuteron. We address also 
the issue of extraction of polarized deep inelastic structure function of the neutron.}

\section{Introduction}
The deuteron is the simplest nuclear system, which is barley bound 
with the rms radius of about $4~fm$. This fact 
in the momentum space is reflected  in the very steep momentum 
distribution of the unpolarized deuteron wave function with 
the strength concentrated predominantly at the small internal 
momenta, Fig.1. However the fact that the 
deuteron has a $D$ wave which vanishes at small momenta 
indicates that isolating the $D$ wave in any 
given nuclear reaction with the deuteron as a target will allow effectively 
suppress the small-momentum/long-range contributions.
This can be seen from the polarized density matrices of the deuteron\cite{FGMSS95}:
\begin{eqnarray}
&&\rho_{d}^{\vec a}(k_1,k_2))    =   u(k_1)u(k_2) +
\nonumber \\
\nonumber \\
&&\left[1 - {3|k_2\cdot a|^2 \over k_2^2} \right]{u(k_1)w(k_2) \over \sqrt{2}} +
\left[1 - {3|k_1\cdot a|^2 \over k_1^2} \right]{u(k_2)w(k_1) \over \sqrt{2}}
\nonumber \\
\nonumber \\
&&+\left( {9\over 2} { (k_1 \cdot a)(k_2\cdot a)^*(k_1\cdot k_2)
 \over k_1^2k_2^2 }-
{3\over2}{|k_1\cdot a|^2\over k_1^2} - {3\over2}{|k_2\cdot a|^2\over k_2^2} +
{1\over 2}
\right) w(k_1)w(k_2),
\nonumber \\
\label{rho_a}
\end{eqnarray}
where $u(k)$ and $w(k)$ represent the $S$ and $D$ partial waves respectively.
The polarization vector  $\vec a$ is defined through the deuteron spin wave
functions:
\begin{equation}
\psi^{10} = i\cdot a_z , \  \psi^{11} = -{i\over \sqrt{2}}(a_x + ia_y) ,
                         \  \psi^{1-1} = {i\over \sqrt{2}}(a_x -ia_y),
\label{VK}
\end{equation}
where $\psi^{1\mu}$ is the projection of the deuteron's spin on the $\mu$
direction. The unpolarized density matrix of the deuteron is defined as:
$\rho^{unp}_d(k_1,k_2)  = {1\over 3} \sum_a  \rho_{d}^{a}(k_1,k_2)$.
\begin{figure}[htb]
\centerline{\epsfxsize=2.4in\epsfbox{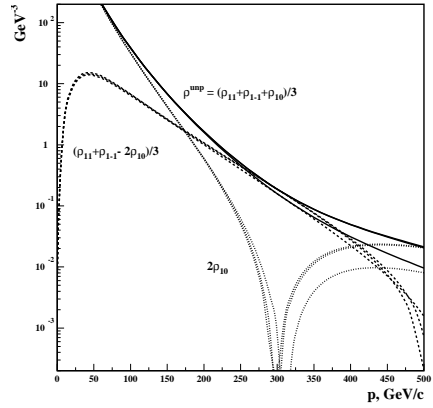}}
\caption{Momentum dependencies of different combination 
of polarized density matrices. Solid, dashed and doted curves correspond 
to  unpolarized, tensor polarized and transverse polarized  distributions 
respectively.}
\label{fig_uva_1}
\end{figure}
Since $\lim_{p\rightarrow 0}w(p) =  0$, it follows from Eq.(\ref{rho_a}) that 
any polarization combination of $\rho_{d}^{\vec a}$,  in  which
$u^2$ term is canceled has an enhanced sensitivity to  the larger 
internal momenta (smaller distances)  of the  deuteron  as compared to the 
unpolarized case.  As it follows from 
Eq.(\ref{rho_a}),  the $u(k_1)u(k_2)$ term does not depend on the polarization 
vector $\vec a$, thus one can cancel this term summing any two components 
of the density  matrix and subtracting the doubled value of the third 
component. 

Fig.\ref{fig_uva_1} presents the  examples of the density matrices for 
unpolarized ($(\rho_{11} + \rho_{1-1} + \rho_{10})/3$,
transverse $\rho_{10}$ and tensor  polarized  
($(\rho_{11} + \rho_{1-1} - 2\rho_{10})/3$ deuteron targets 
as they enter in the impulse approximation term of the  $\vec d(e,e'p)n$ 
cross section (in this case $k_1=k_2=p$). 
As it can be seen from Eq.(\ref{rho_a}) the tensor polarized density matrix
depends only  on the terms proportional to $u(p)w(p)$ and $w(p)^2$ which results in  
a substantial suppression of low-momentum part of the density matrix as compared  to
the unpolarized one .  

We will discuss several studies which utilize  this
unique feature, that choosing special polarization for the deuteron target  
enhances the short-range space-time aspects of the reaction under the 
consideration.

\section{Exclusive $e+ \vec d\rightarrow e'+ p+n$ reaction at $Q^2\ge 1~GeV^2$.}

The density matrices presented in Fig.\ref{fig_uva_1} are probed  
in the electro-disintegration reaction of the polarized deuteron. 
There are two aspects in studies of these reactions. One corresponds 
to the kinematics dominated by impulse approximation (IA) which 
will provide us with the tool for studies of the deuteron properties at 
small inter-nucleon distances. Another aspect corresponds to the kinematics 
dominated by final state interactions (FSI) which is a testing ground  
for studies of the mechanism of FSI at short distances. In general these 
two contributions are intertwined together and  their separation is  an important 
problem.

\begin{figure}[htb]  
\vspace{-0.4cm}  
\centerline{\epsfxsize=2.4in\epsfbox{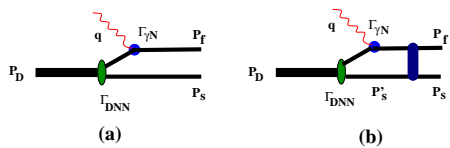}}
\caption{Diagrams for $e+d\rightarrow e'+p+n$ reaction. (a) - IA  and (b) FSI 
         contributions.} 
\label{fig_uva_2}  
\end{figure}

We discuss $d(e,e,p)n$ reaction in the kinematics in which almost 
entire  momentum of the virtual photon is transferred to the knocked-out nucleon 
(for certainty we choose it to be  a proton) with momentum $p_f$, 
while the spectator neutron, with momentum $p_s$ is  detected in the momentum 
range of $0-500$MeV/c.  The main kinematic conditions are: 
\begin{equation}
Q^2 = {\bf q^2} - q_0^2 > 1~ GeV^2, \ \ \ \  {\bf p_f} \approx {\bf q}, \ \ \ \ 
p_f \gg p_m, p_s, 
\label{kin}
\end{equation}
where ${\bf p_m} = {\bf p_f} - {\bf q} = -{\bf p_s}$ is the missing momentum of the 
reaction, $m$ is the mass of the nucleon.
In the kinematic region of Eq.(\ref{kin}) only the IA (Fig.(\ref{fig_uva_2})~(a)) 
and FSI (Fig.(\ref{fig_uva_2})~(b)) terms will dominate the cross section 
(see e.g. Ref.\cite{FSS97,SM01}). The FSI contribution 
in $Q^2\ge 1$~GeV$^2$ limit can be calculated within generalized eikonal 
approximation~(GEA)\cite{FGMSS95,FSS97,SM01} which accounts for the finite 
values of recoil/missing momenta (note that in the Glauber theory the Fermi 
momenta of interacting nucleons are neglected). With IA and FSI terms 
included, the differential cross section of  $\vec d(e,e,p)n$ reaction within 
distorted wave impulse approximation  can be represented as:
 \begin{equation}
{d\sigma\over dE_{e'} d\Omega_{e'} d^3p_f } =
\sigma_{ep}\cdot S_{d}^{\vec {s}}(q,p_f,p_s)\cdot
\delta(q_o - M_d - E_f - E_s),
\label{CRS}
\end{equation}
were  $\sigma_{ep}$  is the cross section of the electron  scattering off a
bound proton (up to the flux and proton recoil factor) and 
$S_{d}^{\vec {s}}(q,p_f,p_s)$ is the distorted spectral function of the deuteron
\cite{footnote}. Within GEA, for $S_{d}^{\vec {s}}(q,p_f,p_s)$ one obtains\cite{FGMSS95}
\begin{eqnarray}
S_{d}^{\vec {s}}(q,p_f,p_s) & = & \rho_d^{\vec s}(p_m,p_m) -
{\mathcal R}e{1\over 2i} \int \rho_{d1}^{\vec s}(p_m,p_m')\cdot f^{np}(k_t)\cdot
{d^2k_t\over (2\pi)^2} \nonumber \\
& & + {1\over 16} \int \rho^{\vec s}_{d2}(p_{m1},p_{m2})\cdot
f^{np}(k_{t1})\cdot
f^{np*}(k_{t2})\cdot
{d^2k_{t1}\over (2\pi)^2} {d^2k_{t2}\over (2\pi)^2},
\label{S_GEA}
\end{eqnarray}
were $\rho^{\vec s}_d$ is defined in Eq.(\ref{rho_a}),  whereas  
$\rho^{\vec s}_{d1}$ and $\rho^{\vec s}_{d2}$ represent the distorted, due 
to the re-scattering, density matrices of the deuteron\cite{FGMSS95,SM01}. $f^{np}$ 
is the amplitude of small angle proton-neutron scattering.

\begin{figure}[htb]  
\vspace{-0.4cm}  
\centerline{\epsfxsize=2.4in\epsfbox{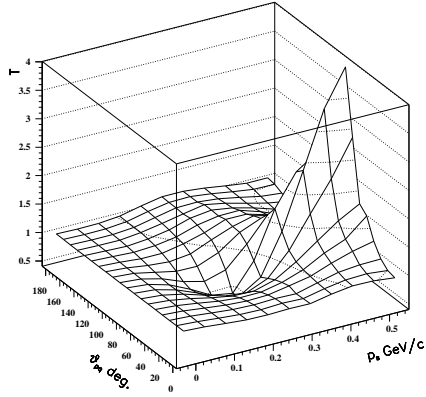}}
\caption{The dependence of the transparency T on the angle - $\theta_{sq}$  
and the  momentum, $p_s$ of the recoil nucleon. The angle is defined with 
respect to the ${\bf q}$.} 
\label{fig_uva_3}  
\end{figure}

To analyze the role of the  IA and FSI in the $d(e,e'p)n$ reaction we 
calculate the ratio of the cross section of Eq.(\ref{CRS}) to the 
cross section calculated within IA only,~(Fig.(\ref{fig_uva_2}(a)):
\begin{equation}
T = {\sigma^{IA+FSI}\over \sigma^{IA}} = {S(p_f,p_s)\over |\psi_D(p_s)|^2}.
\label{T}
\end{equation}
Figure \ref{fig_uva_3}  demonstrates the calculation of $T$ as a function of the 
recoil nucleon angle $\theta_{sq}$ with respect to the ${\bf q}$ for the different 
values of recoil nucleon momentum. It demonstrates the distinctive angular 
dependence of the ratio T. At recoil nucleon momenta $p_s\le 300 MeV/c$,  
$T$ has a minimum and generally  $T<1$ while  at $p_s >300 MeV/c$, $T >1$ and  has a 
distinctive maximum. It can be seen from this  picture that the FSI 
is small at kinematics in which recoil momenta of the reaction is parallel or
anti-parallel to $\bf q$ (referred to as collinear kinematics).
The FSI dominates in the kinematics where  $\theta_{pq}\approx 90^0$, 
more precisely the maximal re-scattering corresponds to the kinematics in which 
$\alpha\equiv {E_s-p_{sz} \over m} = 1$ (referred to as transverse kinematics).
The analysis of Fig.\ref{fig_uva_3} shows that one indeed can isolate 
the kinematic domains where IA term is dominant from the domain in which 
FSI plays a major role. 
The ability to identify these two kinematics is an important advantage of 
$e+d\rightarrow e'+p+n$ reactions. It allows to concentrate on the different aspects of 
the dynamics of $d(ee'p)n$ reaction with less background effects. Namely the collinear
kinematics are best suited for studies of bound nucleon dynamics while in transverse 
kinematics one can concentrate on the physics of hadronic re-interaction.

\subsubsection{Studies of electro-production from deeply bound 
nucleons, relativistic effects}

The $d(e,e'p)n$ reaction  within IA, Fig.~\ref{fig_uva_2}(a) represents the 
testing ground for investigation of the electromagnetic structure of bound nucleons. 
Within IA there is a direct correspondence between the momentum of spectator neutron and 
the binding energy of the interacting proton $E_b = E_m = m_d-\sqrt{m^2+p_s^2}$. To 
achieve such a simplicity in studies of bound nucleon structure, according to 
the discussions in the previous section, one has to choose collinear kinematics 
in which FSI is a correction.  Additionally, one should restrict the $Q^2\le 4$ GeV$^2$
in which  case we expect minimal color transparency effects (see next section) and 
therefore the FSI, being small, are also  well under the control.  
Furthermore, we focus on the parallel kinematics in which the light cone momentum 
of spectator nucleon, $\alpha\equiv {E_s-p_s^z\over m} > 1$ and $p_\perp\approx 0$.
The $d(e.e'p)n$ reaction in these kinematics are most sensitive to relativistic effects 
in the deuteron\cite{rep}. The sensitivity to relativistic effects persists also 
at  angles of spectator nucleon momenta close to the  collinear kinematics. This sensitivity 
gradually disappears at $\alpha\rightarrow 1$ and $\theta_{sq}\rightarrow 90 ^0$. 
There are several techniques for treating the deeply bound 
nucleons as well as relativistic effects in the deuteron. One group of approaches 
handles the virtuality of the bound nucleon within a description of the deuteron in the lab. frame 
(we will call them virtual nucleon (VN) approaches)  by taking the residue over the energy of the 
spectator nucleon. One has to deal with negative energy states which arise  for 
non-zero virtualities (see e.g. Ref.\cite{MST}).  Due to the binding, the current 
conservation is not automatic and one has to introduce a prescription to 
implement electromagnetic gauge invariance (see e.g. Ref.\cite{deF}). 

Another approach is based on the observation that high energy processes
evolve along the light-cone.  Therefore, it is natural to describe the 
reaction within the light-cone~(LC) non-covariant framework \cite{rep}. 
Negative energy states do not enter in this case, though one has to take into 
account so called instantaneous interactions.
For this purpose one employs e.m. gauge invariance to 
express the ``bad'' electromagnetic current component 
(containing instantaneous terms) through the ``good'' component 
$J^A_+ = -q_+/q_-J^A_-$ \cite{rep}.
In the approximation when non-nucleonic degrees of freedom in the
deuteron wave function can be neglected, one can unambiguously relate
the light-cone wave functions to those calculated in the Lab frame
by introducing the LC $pn$ relative three momentum
$k=\sqrt{{m^2+p_t^2\over \alpha(2-\alpha)} - m^2}$.

\begin{figure}[htb]
\centerline{\epsfxsize=2.4in\epsfbox{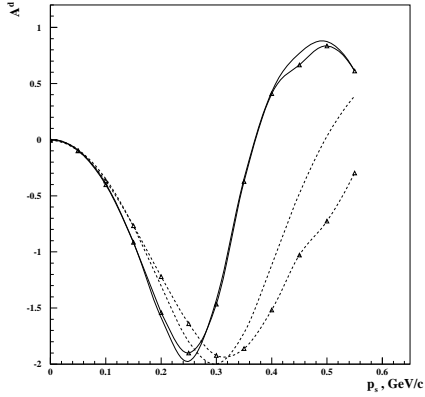}}
\caption{$p_s$ dependence of the $d(e,e'p)n$
tensor polarization asymmetry at $\theta_{sq}=180^0$. Solid
and dashed lines are IA predictions
of the LC and VN methods, respective
marked curves include FSI.}
\label{fig_uva_4}
\end{figure}

Naturally, VN and LC approaches coincide in  the limit of small missing momenta.
Their predictions within IA considerably diverge at larger values of 
spectator momenta $(\ge 300 MeV/c$)\cite{rep}. 
According to the discussion  in  Sec.1 the  $d(e,e'p)n$ reactions
with a tensor polarized deuteron is best suited for discrimination between 
VN and LC  prescriptions. This was previously demonstrated within IA in Ref.\cite{rep}. 
Using the recent advances in the calculation of FSI, 
one can perform a similar comparison for  asymmetry:
\begin{equation}
A^{d} = {\sigma^{1,1} + \sigma^{1,-1} - 2 \sigma^{1,0}\over \sigma^{1,1} + \sigma^{1,-1} + \sigma^{1,0}},
\label{Ad}
\end{equation}
accounting also for FSI contribution of Fig.\ref{fig_uva_2}(b). 
In Eq.(\ref{Ad}) $\sigma^{1,s_z}\equiv {d\sigma^{\vec s,s_z}\over dE_{e'} d\Omega_{e'} d^3p_s}$ 
represents the differential cross section of $d(e,e'p)n$ reaction with 
the deuteron helicity, $s_z$.

The results of VN and LC comparison which includes both IA and FSI contributions are presented 
in  Fig.\ref{fig_uva_4} for collinear kinematics with $\theta_{sq}=180^o$. One can see that 
account of the FSI further increases the difference between VN and LC predictions, thus making 
their experimental investigation more feasible.

\subsubsection{Color Transparency studies at intermediate $Q^2$}

In QCD, the absorption of a high $Q^2$ photon by a nucleon produces a 
point like configuration~(PLC), which, at asymptotically high energies, would not 
interact with the nucleons, thus eliminating FSI.  This effect generally is 
referred as a color transparency (CT). Recently, CT was experimentally observed\cite{Ashery} in the 
high energy $\pi + A \rightarrow 2jets + A^\prime$ reactions which confirmed the early prediction
based on perturbative QCD (pQCD) calculations\cite{FMS}.

At high but finite energies (pre pQCD domain), a PLC is actually produced, but it  
expands as it propagates through the nucleus \cite{ANN}.
To suppress the expansion effects, it is necessary to ensure that 
the expansion length, $l_h\sim 0.4(p/$GeV), is greater than the
characteristic longitudinal distance in  the reaction.
In the considered $d(e,e'p)n$ reaction, where one nucleon
carries almost all the momentum of the photon while the second nucleon 
(or its resonance) is a spectator, the actual expansion
distances are the distances between the nucleons in the
deuteron \cite{FGMSS95}. Thus, suppressing large distance effects
through the deuteron's {\em polarization}, one effectively will diminish the 
PLC's expansion, leading to an earlier onset of CT.

\begin{figure}[htb]
\centerline{\epsfxsize=2.4in\epsfbox{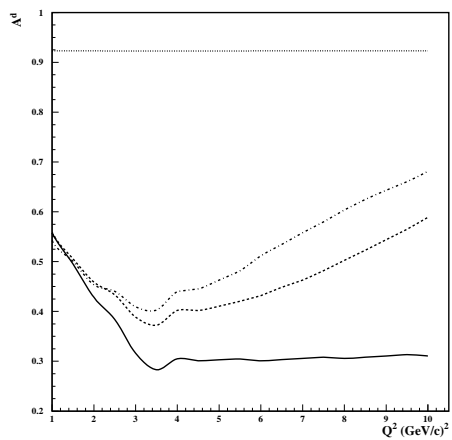}}
\caption{$Q^2$ dependence of $A_d$ for $\alpha =1$. 
Solid line - GEA, dashed - QDM,
dashed-dotted - three state model, dotted -PWIA.}
\label{fig_uva_5}
\end{figure}

The reduced interaction  between the PLC and the spectator nucleon can be described 
in terms of its transverse size and the distance $z$ from the 
photon absorption point in $d(e,e'p)n$ reaction. As a result the $pn$ re-scattering 
amplitude, $f^{pn}$, in Eq.(\ref{S_GEA}) is replaced by
$f^{PLC,N}(z,k_t,Q^2)$. For numerical estimates of the $f^{PLC,N}(z,k_t,Q^2)$,
we use the quantum diffusion model (QDM) \cite{FLFS} as well as the three 
state model \cite{FGMS93}. Latter is based on the assumption 
that the hard scattering operator acts on a nucleon and produces a  
PLC, which is represented as a superposition of three baryonic states, 
$|PLC\rangle = \sum_{m=N,N^*,N^{**}} F_{m,N}(Q^2) |m\rangle$.

To study the expected effect of CT we choose now a transverse kinematics, 
in which FSI is dominant (see Fig.\ref{fig_uva_3} . Furthermore we choose the 
tensor polarization  of the deuteron target which is sensitive to the $D$ wave  and 
therefore FSI will be dominated at smaller inter-nucleon distances.  
As a result one will probe the evolution  
of PLC at  smaller space-time intervals. 

For numerical estimates, we consider the $Q^2$ dependence of the asymmetry $A_d$ 
from  Eq(\ref{Ad}) for fixed and transverse momenta of the spectator neutron.
This dependence for $p_t=300$ MeV/c, is presented in Fig.\ref{fig_uva_5} 
One can see from this figure that CT effects  
can change $A_d$ by as much as factor of two for $Q^2\sim 10$ GeV$^2$.
It is worth noting that the same models predict 15-20\% effect for 
$(e,e'p)$ reactions on unpolarized nuclear targets.

\section{Coherent production of  vector mesons from polarized deuterons at high $Q^2$}

The ability to select smaller than average inter-nucleon distances in 
the reactions involving polarized deuteron target is explored further in 
the reactions of coherent electro-production of vector mesons, i.e. 
$e+\vec d\rightarrow e'+d'+V$, in which $V$ represents a neutral mesons i.e. 
$\rho$, $\omega$, $\phi$ etc.
\begin{figure}[htb]
\centerline{\epsfxsize=2.4in\epsfbox{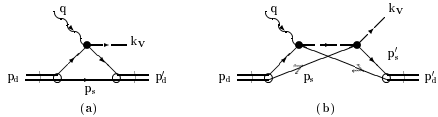}}
\caption{IA (a) and FSI (b)  contributions to $d(e,e'V)d'$ reaction.}
\label{fig_uva_6}
\end{figure}

In this case two, impulse approximation~(IA) and single re-scattering  diagrams define 
the overall cross section (see Fig.\ref{fig_uva_6}).  At high $Q^2$ one may expect that 
the intermediate state produced by the virtual photon in the FSI amplitude 
(Fig.\ref{fig_uva_6}(b)) will be dominated by PLCs. As a result one may observe a 
diminished re-interaction with the spectator nucleon. The philosophy here is similar 
to that of deuteron electro-disintegration reaction, which is too look for specific 
polarization of target deuteron in order to achieve the highest possible sensitivity to 
the FSI term contributing to the cross section of the reaction. 

To see the role of the deuteron polarization we analyze the production amplitude of 
$\gamma^*+d\rightarrow d'+V$ reaction, which be written as follows\cite{vmeson1}
\begin{eqnarray} \label{eq:amplitude} 
F_{df}^{j j'} &=& F^{(a)} + F^{(b)} =   
f^{\gamma^*N\rightarrow VN}({\bf l})
\left[
S^{j j'}_{df}\left(-{{\bf l_{\perp}}\over 2}, 
{l_{\_} \over 2}\right)~+~
S^{j j'}_{df}\left({{\bf l_{\perp}}\over 2}, 
-{l_{\_} \over 2}\right)\right] 
\nonumber \\ 
&+&
{i\over 2}\sum_h \int 
{d^2 k_\perp  \over  (2\pi)^2}
f^{\gamma^*N\rightarrow hN}
\left({{\bf l_\perp}\over 2}-{\bf k_\perp}\right) 
f^{hN\rightarrow VN} 
\left({{\bf l_\perp}\over 2}+{\bf k_\perp}\right) \nonumber \\
&&\hspace{1.5cm}\times
\left[
S^{j j'}_{df}({\bf k_\perp},-\Delta_h) + 
\frac{2 i}{\sqrt{2\pi}} 
\Delta S^{j j'}_{df}({\bf k_\perp},-\Delta_h)
\right].
\end{eqnarray}
where $S^{j j'}_{df}$ is the transition form-factor of the deuteron, 
$f^{\gamma^*N\rightarrow VN}$  and $f^{hN\rightarrow VN}$ corresponds to the 
amplitudes of $\gamma^*+N\rightarrow N+h$ and $h+N\rightarrow N+V$ reactions 
respectively. $l_- = l_0 - l_z$, where $l_0$ and $\bf l$ are the transferred energy 
and momentum to the final deuteron.
The transition form-factor of the deuteron, $S$ forms the density matrix that 
enters in  the differential cross section of the reaction. In the general form 
this density matrix can be represented as follows:
\begin{eqnarray} \label{eq:density_matrix_explicit}
& & \rho^{1, m}({\bf l_1},{\bf l_2}) = 
\sum_{m'} S_d^{s,m m'}({\bf l_1}) S_d^{1,m m'}({\bf l_2})^{\dag} 
=  F_C(l_1) F_C(l_2)  
\nonumber \\
& & + 
\frac{1}{\sqrt{2}} \left\{ \left[
\frac{3|{\bf l_2} \cdot {\bf \epsilon_1^m}|^2}{l_2^2} -1
\right] F_C(l_1) F_Q(l_2) + \left[
\frac{3|{\bf l_1} \cdot {\bf \epsilon_1^m}|^2}{l_1^2} -1
\right] F_C(l_2) F_Q(l_1)  \right. \nonumber \\
& & + 
\left.\left[9\frac{({\bf l_1}\cdot {\bf \epsilon_1^m})
            ( {\bf l_1}\cdot {\bf \epsilon_1^m})^* 
              {\bf l_1}\cdot {\bf l_2}}{l_1^2 \,l_2^2} 
- \frac{3|{\bf l_1} \cdot {\bf \epsilon_1^m}|^2}{l_1^2} 
- \frac{3|{\bf l_2} \cdot {\bf \epsilon_1^m}|^2}{l_2^2}  + 1 \right]\right.
\nonumber \\
& & \ \ \ \ \ \ \ \ \ \ \  \ \ \ \ \ \ \ \ \ \ \ \ \ \ \ \ \ \ \ \ \ \  \ \ \ \ \ \ \ \ \ \ \
\ \ \ \ \ \ \ \ \ \ \ \ \  \ \ \ \ \ \ \ \ \ \ 
\left. \times \frac{F_Q(l_1) F_Q(l_2)}{2}\right\},
\label{den} 
\end{eqnarray}
where $F_C$ and $F_Q$ are the charge and quadrupole form-factors of the deuteron and 
${\bf \epsilon_1^m}$ is the polarization vector of the target 
deuteron with $m$ being the deuteron spin projection on the given quantization axis.

\begin{figure}[htb]
\centerline{\epsfxsize=2.4in\epsfbox{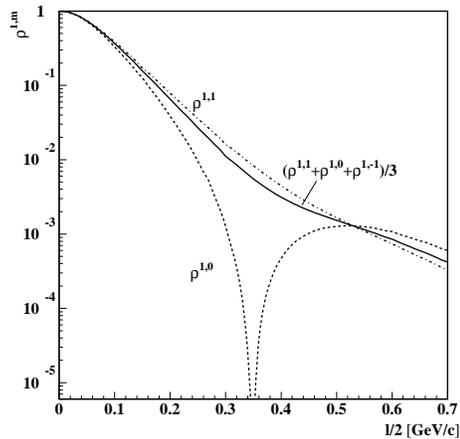}}
\caption{The deuteron density matrix $\rho^{1,m}$ for a spin quantization 
axis parallel to the photon momentum $\bf q$. The dashed and dash-dotted 
lines correspond  to spin projections $m=0$ and $m=1$, respectively.
The density matrix for an unpolarized  deuteron is shown by 
the solid curve.}
\label{fig_uva_7}
\end{figure}

In the IA term the $\rho$ function enters with the argument ${\bf l_1}={\bf l_2}={\bf l/2}$.
Since our aim is to identify the kinematic regions in which the IA term vanishes and FSI is dominant, 
we search for those polarizations of the deuteron for which 
the $\rho^{1, m}({{\bf l}\over 2})$ has a vanishing values for accessible range of ${\bf l}$.
The analysis of the density matrix as it enters in the IA term of the cross section is 
presented in Fig.\ref{fig_uva_7}. The figure demonstrates that for transversely polarized 
deuterons the $\rho^{1, 0}({{\bf l}\over 2})$ vanishes  at $l\approx 0.7$~GeV/c which corresponds to 
$-t\approx0.5$~GeV$^2$. The existence of the zero in $\rho^{1,m}({{\bf l}\over 2})$ can be understood
in the limit of vanishing $l_-$, in which case 
 $\rho^{s, 1}({{\bf l_\perp}\over 2}) = (F_C(l_\perp/2) - {1\over \sqrt{2}}F_Q(l_\perp/2))^2$. 
Since $F_c$ monotonically decreases while $F_Q$ increases from the zero value 
at $l_\perp =0$, the following  equation has a solution $F_C(l_\perp/2) = {1\over \sqrt{2}}F_Q(l_\perp/2))$ 
which  is found to be at $l_\perp\approx 0.7$~GeV/c .
Thus our observation is that one has vanishing IA contribution 
for $d(e,e'V)d'$ reaction for transversely polarized deuteron target at $-t\approx 0.5$~GeV$^2$. 
Therefore one expects that at these $t$ the cross section should exhibit strong sensitivity to 
the dynamics of re-interaction of $h$ with the spectator nucleon. 
This situation gives us a clue on where to look for  the CT effects that reveal themselves 
through the decrease of FSI contribution~(Fig.\ref{fig_uva_6}(b) with an increase of $Q^2$.

\begin{figure}[htb]
\centerline{\epsfxsize=2.4in\epsfbox{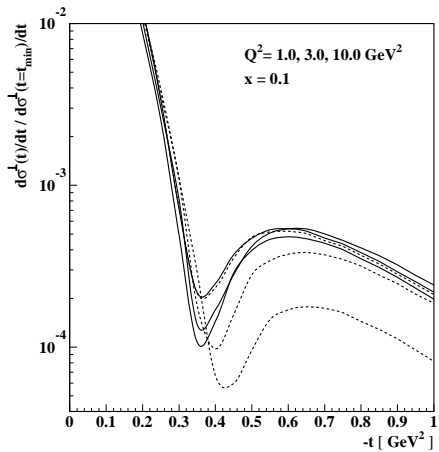}}
\caption{ The cross section $d\sigma_{\gamma^*N}^{\perp}/dt$ normalized 
to its value at $t=t_{min}$. The solid lines present the complete vector 
meson dominance calculation.  Results from quantum diffusion are  shown by 
the dashed curves. The cross sections decrease with an increase of $Q^2$.}
\label{fig_uva_8}
\end{figure}

In Fig.\ref{fig_uva_8} we estimate the expected CT effects for different values of 
$Q^2$ based on QDM model of CT\cite{FLFS} discussed in the previous section.  
As Fig.\ref{fig_uva_8} demonstrates 
the onset of CT will result in  qualitative changes in the $t$ dependence of 
the cross section with minimum at $-t\approx 0.45-0.5$~GeV$^2$ which becomes 
increasingly pronounced with an increase of  $Q^2$.

Note that the very same method of isolating 
FSI contribution in $d(e,e'V)d'$ reaction  at $Q^2\approx 0$ 
can be used to determine the interaction cross sections of  vector mesons 
with nucleons.  This is especially important for $\phi$ and $\psi$ mesons 
whose interaction cross section with the nucleon is poorly known\cite{vmeson2}.

\section{Measurement of Polarized DIS Structure Function of the Neutron}

So far we were interested in isolating the $D$ wave contribution 
of the deuteron wave function which allowed to increase the  sensitivity 
to the short distance phenomena in the reaction involving polarized 
deuteron. Now we concentrate on  the separation of $S$ wave contribution 
which will give a possibility to estimate the polarized structure function 
of the ``free'' neutron. Since the deuteron has a spin 1, in the $S$ state one 
has simplified relation between  helicity of the deuteron and the bound neutron. 
To isolate these neutrons in almost free state one considers 
semi-inclusive $e+d\rightarrow e' + p + X$ reaction in which the protons 
are detected in the target fragmentation region. 
Such reaction would be very natural for the electron ion collider in 
the $e \vec{D}$   mode.
By selecting only the slowest recoil protons one should be able to isolate 
the situation whereby the virtual photon scatters from a nearly on-shell 
neutron in the deuteron.   
In this way one may hope to extract the DIS structure functions of the  neutron 
with a minimum of uncertainties arising from modeling 
nuclear effects in the deuteron.  As an example we consider the 
extraction of $g_1^n$ function from $\vec d(\vec e, e'p)X$ reaction in which 
the deuteron is polarized in the direction opposite to the direction of incoming 
electron. In this case the asymmetry with respect to the 
helicity of incoming electron in the Bjorken limit can be expressed as
follows:
\begin{equation} 
\sigma^{\uparrow} - \sigma^{\downarrow} 
\approx  {2\alpha_{em}^2 \over m x (2-\alpha) Q^4} (1-y)\cdot    
 g^\mathrm{{eff}}_{1n}\left({x\over 2-\alpha},Q^2\right)\cdot
[u(p)^2-{w(p)^2\over 2}]
\label{2N} 
\end{equation}  
where $\sigma^{\uparrow (\downarrow)}\equiv {d\sigma^{\uparrow (\downarrow)} \over 
d\phi dx dQ^2 d(\log\alpha) d^2p_\perp }$  in which $\uparrow (\downarrow)$ corresponds to 
positive (negative) helicities of incoming electron. For the asymmetry, in Eq.(\ref{2N}, 
the relativistic corrections due to the Fermi motion are known to be small 
up to $x\sim 0.5$. In this region the deuteron to a very good approximation 
will reduce the asymmetry by the depolarization factor of $1-3/2 P_D$ \cite{FS83}, 
where $P_D$ is the total probability of the $D$ state in the deuteron. The example of 
relativistic description of the density matrix in calculation of $g_1d$ is given in 
Refs.\cite{FS83,TeIs}. It is worth noting that the 
measurement of $g_{1n}$ using inclusive scattering off the deuteron has
certain disadvantages - one has to subtract $g_{1p}$ which is larger in a wide
$x$ range, and there is a question of the EMC type effects. Hence to extract the free $g_{1n}$ 
in a model independent way one would tag spectator protons (a rather easy task for the collider
kinematics) and perform  the measurements at $p\to 0$

To extract the free $g_{1n}$ from Eq.(\ref{2N}) one first measures the limit of $p\rightarrow 0$. 
Since the $D$ wave vanishes in this limit we are sensitive only to the $S$ partial wave. 
Furthermore,  we can  extrapolate the  measured tagged neutron structure  $g^\mathrm{{eff}}_{1n}$  
to the region of negative values of kinetic energy of the spectator proton\cite{hnm}. This method  
is analogous to the Chew--Low procedure for extraction of the cross  
section of scattering off a pion\cite{CL59}. Such an extrapolation allows us to 
isolate  the pole in the $S$ wave which corresponds to  
the on-shell neutron interacting with virtual photon. This pole in the IA amplitude  
is located  at $E_{kin}^{pole}=-{|\epsilon_{D}|-(m_{n}-m_{p})\over 2}$. 
The advantage of this approach is in the fact that the  scattering amplitudes 
containing final state interactions do not have singularities 
corresponding to on-shell neutron states. Thus, isolating the singularities 
through the extrapolation of effective structure functions into the negative 
spectator kinetic energy range  will suppress the FSI effects in the extraction of 
the free DIS structure function.

\section{Outlook}
Use of the polarized deuteron targets in high energy electro-production reaction
gives unique possibility to study the strong interaction at short space-time distances.
These studies are the part of the broad program dedicated to the investigation 
of the structure of deeply bound nucleon as well as the physics of color transparency\cite{hnm}.
In semi-inclusive DIS reactions with low momenta of recoil protons the use of the 
polarized deuterons allow to perform very accurate measurements of polarized structure 
functions of the neutron. This program could benefit tremendously from the advances 
of the building the polarized deuteron targets that can be operated under the high 
current electron beams as well as from building the polarized  electron -ion collider.


\begin{thebibliography}{02}
\bibitem{footnote}In plane wave impulse approximation, in which  
         only the diagram of Fig.(\ref{fig_uva_2})(a) is included, 
         the $S_{d}^{\vec {s}}(q,p_p,p_n)$ represents the 
         probability for the proton in the polarized deuteron to  have a 
         momentum - $p_m$.
\end{thebibliography}
\end{document}